\documentclass[12pt]{article}
\usepackage{amssymb,amsmath,epsfig,cite}
 \allowdisplaybreaks
\begin{document}
\title{\bf Dissipative Self-Gravitating Systems in Modified Gravity}
\author{M. Z. Bhatti$^1$\thanks{mzaeem.math@pu.edu.pk}, Kazuharu Bamba$^2$
\thanks{bamba@sss.fukushima-u.ac.jp}, Z. Yousaf$^1$\thanks{zeeshan.math@pu.edu.pk} and M. Nawaz$^1$\thanks{mubeen.nawaz9443@gmail.com}\\
$^{1}$ Department of Mathematics, University of the Punjab,\\
Quaid-i-Azam Campus, Lahore-54590, Pakistan\\
$^2$ Division of Human Support System,\\ Faculty of Symbiotic Systems Science,\\ Fukushima University, Fukushima 960-1296, Japan}

\date{}

\maketitle
\begin{abstract}
We discuss the gravitational collapse of spherical compact objects in the background of $f(R,T,Q)$ theory, where $R$ represent the Ricci scalar, $T$ is the trace of energy momentum tensor while $Q\equiv R_{\mu\nu}T^{\mu\nu}$, and investigate the influence of anisotropy and heat dissipation in this scenario.
We provide an analysis on the role of distinct material terms considered
while studying the dynamical equation. The
dynamical equation is coupled with a heat transport equation and discussed
in the background of $f(R,T,Q)$ theory of gravity. The reduction element in the density of inertial mass, is re-acquired which is based on the internal position of thermodynamics.
In collation with the equivalence relation, the reduction quantity in density is
similar as appeared with gravitational force. We formulate the connection of Weyl tensor with different matter variables to see the
non-identical outcomes. The inhomogeneous nature of energy density
is also analyzed in the framework of modified gravity.
\end{abstract}
{\bf Keywords:} Self-gravitation; Anisotropic fluid; Gravitational collapse.\\

\section{Introduction}

General theory of relativity (GR) is considered as the foundation of
gravitational physics and cosmology. The significance of its equations
is the important study of how the space time is effected
regardless of concerned matter and radiation. Relevance effects of gravity
such as gravitational waves, gravitational lensing, and
time dilation are also the consequences of GR. In modern astrophysics \cite{1,2}, GR and the big bang
model of cosmology not only depict the movement of the planets,
but also describe the history and extension of the universe, the
phenomena behind black holes and the angle of light from faraway stars and
galaxies.

Gravitational collapse occurs when the astronomical objects are
contracted under the extensive effect of gravity. It draws substance
towards the central point of gravity and has significance to understand the
some characteristic features of our cosmos.
The initial work on gravitational collapse was started in
$1939$ for spherically symmetric stellar objects with dust fluid. Afterwards,
Oppenheimer and Snyder \cite{os1} explained the central effect of collapsing stars
by considering perfect fluid configuration in the stellar interior.
During the last decade, the dynamics of the expanding universe has been
scrutinized under the effects of modified theories.
The straightforward generalization of GR among these
is $f(R)$ theory, obtained by inserting a generic function of the
Ricci scalar in the Einstein-Hilbert action to get some viable results about the cosmos \cite{3,4,5}.
It is a family of theories, determined by non identical function $f$ of the Ricci scalar.
The $f(R)$ gravity is also
suitable to study the cosmic acceleration and the state of feasible
cosmological models (for recent reviews on modified gravity theories as well as the issue of
dark energy, see, e.g.,~\cite{Nojiri:2010wj, Capozziello:2011et, Capozziello:2010zz, Bamba:2015uma, Cai:2015emx, R9, R10, Nojiri:2017ncd, Bamba:2012cp}). The $f(R,T)$ and $f(R,T,Q)$ are the
generalizations of $f(R)$ gravity where $T\equiv g^{\gamma\delta}T_{\gamma\delta}$ and
$Q$ specifies the effects of geometry-matter non-minimal interaction.
Haghani \emph{et al.} \cite{7} derived the equations of motion
in $f(R,T,Q)$ gravity for conservative as well as the non-conservative
system. Odintsov and S\'{a}ez-G\'{o}mez \cite{8} find a way
in $f(R,T,Q)$ theory having non-minimal coupling in
gravitational field and with matter. Ayuso \emph{et al.} \cite{9} examined the
stability condition of $f(R,T,Q)$ with distinction of scalar term as
well as vector field to consider the generalized form of
Einstein-Hilbert action.

Baffou \emph{et al.} \cite{10} studied the advanced evolutionary cosmic phases in the presence
of additional curvature terms due to higher order theory. Elizalde and Vacaru \cite{10a} explored few approximate and exact cosmological models and concluded that the presence of off-diagonal non-vacuum/vacuum distributions in GR could provide interesting theoretical insights of $f(R, T, Q)$ gravity. Yousaf \emph{et al.} \cite{17}
examined the stability conditions for spherical collapse with the
contribution of $f(R,T,Q)$ extra curvature ingredients by imposing perturbation
and concluded that extra curvature terms slows down the collapse rate.

Chan \emph{et al.} \cite{11,12} analyzed pressure anisotropic effects
on the existence and evolution of compact objects. Herrera \emph{et al.} \cite{13}
investigated the evolution of collapsing structures by taking zero value to the expansion scalar. They found that
their constraints of dynamical instability are independent of stiffness parameter due to the presence of vacuum core.
The study of dynamical instability constraints for non-static
compact objects when coupled with the imperfect fluid in
astrophysics and cosmology are being discussed by various researchers \cite{14,15,16}.
They concluded that the dark terms of curvature tend to increase the regions of stability.
The role of expansion scalar in the investigation of instability regions for a stellar interiors
has been analyzed by \cite{16a}. Herrera \cite{16b} analyzed some dynamical features of the radiating gravitational
collapse via transport and dynamical equations. Herrera
\emph{et al.} \cite{16c} studied modeling and evolution of anisotropic stellar collapse.

It is an evident fact that gravitational collapse is purely a dissipative phenomena, so the importance of its
effects in the description of self-gravitating objects cannot be ignored. Dissipation can be well-discussed in the background of two approximations, i.e., free streaming and diffusion. In the latter case, the heat radiation is directly proportional to the inclination of its temperature \cite{18}. It indicates the passage of the particle as a primary cause of energy proliferation in the interior of celestial objects which are typically microscopic in size. Therefore, a main sequence star like the sun becomes very small in size as compared to the stellar core. Yousaf \cite{somep} studied the dynamical properties of self-gravitating systems and calculated few exact analytical solutions. He also calculated inhomogeneity factors with the help of Weyl tensor. Stabile and Capozziello \cite{sta1} studied the limits of weak approximation for the Einstein and Jordan for the self-gravitating systems and presented the transformation scenario of structural effective variables from one frame to another.

The dissipation effects for the radiative transport in both
limitations have been discussed by Herrera and Santos \cite{19} in accordance to quasi-static
approximation. The use of this estimation is quite reasonable because
the measurement of hydrostatic time is much less in comparison to
star's lifespan, for the various stages of star. For
the sun it is, $27$ min, in case of white dwarf it is $4.5$ sec, and it
becomes $10^{-4}$ for neutron star having same solar mass with same radius
of $10$ km. Nevertheless, it is difficult to apply this estimation to
all dynamic stages that was discussed earlier, where the complete
dynamic association had to be used and it is required to consider
all the conditions which relate the state of leaving equilibrium.

Here, the focus of discussion is to examine the behavior of $f(R,T,Q)$ gravity
for anisotropic spherically symmetric stellar objects undergoing collapse in the
presence of heat dissipation \cite{20}. The format of the paper is such a way that the coming section, we will explore
the kinematical variables for spherically symmetric stars and the field equations in the background of $f(R,T,Q)$ theory. Section \textbf{3} deals with conservation equations to construct the dynamical equation. In section \textbf{4}, we explore the heat transport equations and couple it with the dynamical equation to study the process of gravitation collapse. In section \textbf{5}, we exhibit an explicit relation of the Weyl tensor with matter variables in the framework of modified theory. The last section summarizes the whole discussion.

\section{Stress Energy Tensor and $f(R,T,Q)$ Field Equations}

In this section, we will provide a systematic study of spherical geometry coupled with anisotropic radiating fluid.
The kinematical quantities and modified field equations are explored to continue our analysis.

\subsection{Interior Spacetime}

We consider an anisotropic radiating matter distribution, around the area covered with
spherical surface $\Sigma$. We assume that the fluid is accompanied
with dissipation effects under diffusion approximation with comoving coordinates.
The general form of the interior metric is given by
\begin{equation}\label{1}
ds^2_{-}=B^2dr^{2}+C^2(d\theta^{2}+\sin^2\theta{d\phi^2})-A^2dt^{2}.
\end{equation}
Inside the boundary surface $\Sigma$, the energy momentum tensor for the anisotropic and dissipative matter configurations in the background of spherical geometry is given by
\begin{equation}\label{2}
T^{-}_{\alpha\beta}=(P_{\bot}+\mu)V_{\alpha}V_{\beta}+q_{\alpha}V_{\beta}
+\chi_\alpha\chi_\beta(P_{\bot}-P_r)+P_{\bot}g_{\alpha\beta}+V_{\alpha}q_{\beta},
\end{equation}
where where $\mu,P_\perp,~P_r,~q_\alpha$ denote the energy density, the tangential pressure,
the radial pressure, the heat flux vector describing dissipation at the diffusion
approximation, respectively. The four-velocity of any fluid element $(V_\alpha)$ is time-like. The quantity $\chi^\alpha$ is a space-like vector directed along the radial direction. All these vectors under comoving system obey
\begin{equation}\label{3}
V_{\alpha}V^{\alpha}=-1, V^{\alpha}q_{\alpha}=0,\quad
\chi^{\alpha}\chi_{\alpha}=1,
\end{equation}
relations. There has been very interesting works of the evolution of relativistic systems
on non-comoving (tilted) frames \cite{hernew1}.

Herrera and Santos \cite{7g} discussed that the gravitational collapse is highly dissipative phenomenon to study the physical characteristics of stellar objects. Ruderman \cite{z1g}, Cameron and Canuta \cite{z2g} proposed different approaches which include certain modeling of local anisotropic fluid distribution while studying the gravitational collapse rate. Many researchers regarded anisotropy as collision-less occurrence while observing its implications on collapse such as, Michie \cite{z3g}, Binney and Tremaine \cite{z4g}. Under some specific configuration of dark matter in galactic haloes, the concerned matter distribution or angular momentum during the formation of spherically symmetric objects remain conserved which lead to emergence of anisotropy.

Certainly, the formation and evolution of self-gravitating systems depend upon the anisotropy, specific thermal condition and the heat radiation of the matter distribution. The matter content of this paper represents the anisotropic behavior of the fluid distribution undergoing diffusion approximation which cause radiations in the form of heat flux as represented with $q_{\alpha}$. The shear tensor can be characterized in the general form
\begin{equation}\label{4}
\sigma_{\gamma\delta}=V_{({\gamma;\delta})}+a_{(\gamma} V_{\delta)}-\frac{1}{3}\Theta(g_{\gamma\delta}+V_{\gamma}V_{\delta}),
\end{equation}
where $a_{\gamma}$ (four acceleration) and $\Theta$ (expansion scalar) can be given as follows
\begin{equation}\label{5}
a_{\gamma}=V_{\gamma;\mu}V^{\mu},\quad \Theta=V^{\gamma}_ {;\gamma}.
\end{equation}

To evaluate the quantities ${\mu},~P_{r},~P_{\bot},~q^{\alpha}$, we need particular relations linked with
these quantities, for a significant physical model.
It should be noted that for
some general significant results, which is our assumption,
dissipation takes energy radially in the outward direction.
For our spherically symmetric line element, the unit vectors in
comoving coordinates can be taken as
\begin{equation}\label{6}
V^{\alpha}=A^{-1}\delta^{\alpha}_{0},\quad
q^{\alpha}=qB^{-1}\delta^{\alpha}_{1},\quad
\chi^{\alpha}=B^{-1}\delta^{\alpha}_{1},
\end{equation}
here $q$ depends on $t$ and $r$. By utilizing (\ref{6}), we
explore the non-null components of shear tensor as
\begin{equation}\label{7}
\sigma_{11}=\frac{2}{3}B^{2}\sigma,\quad
\sigma_{22}=\frac{\sigma_{33}}{\sin^{2}\theta}=-\frac{1}{3}(C)^{2}\sigma,
\end{equation}
where
\begin{equation}\label{8}
\sigma=-\left(\frac{\dot{C}}{AC}-\frac{\dot{B}}{AB}\right),
\end{equation}
where, $.\equiv\partial/\partial t$.
In a similar fashion, we obtain the values of four acceleration and expansion scalar
from Eqs.(\ref{5}) and (\ref{6}) as
\begin{equation}\label{9}
a_{1}=\frac{{A'}}{A},
\Theta=\frac{1}{A}\left(\frac{\dot{B}}{B}+2\frac{\dot{C}}{C}\right),
\end{equation}
where $'\equiv\partial/\partial r$

\subsection{$f(R,T,Q)$ Equations of Motion}

In this section, we will discuss the exploration of field equation in a particular
modified gravity namely, $f(R,T,Q)$ gravity. The formulation of this gravity theory
is based on the input of non-minimal coupling (NMC) between matter and geometry. The basic structure of Einstein-Hilbert action is modified in order to elucidate the role of dark matter and dark energy, without resorting to some exotic fluid configurations. Initially, it was focused on the modification of geometric part of the action via substitution of the Ricci scalar with its generic function $f(R)$. Though there are many $f(R)$ theories that
can incorporate the chameleon mechanism compatible with solar system constraints. \cite{f10}. Consequently, a general coupling between matter Lagrangian ($L_m$) and Ricci scalar as a maximal extension of the Einstein Hilbert action was established by Harko and Lobo \cite{1g}, where the Lagrangian is a generic function of $R$ and $L_m$, and it termed as $f(R,L_m)$ theory of gravity. The non-minimal coupling between matter and geometry in $f(R,L_m)$ gravity has driven an additional force as described by \cite{f23}. This force characterized the point test particle to behave non-geodesically.

By utilizing the basic concepts studied by Poplawski \cite{2g}, the general NMC between matter and geometry was considered in the framework of Lagrangian of the form $f(R,T)$, containing an arbitrary function of the Ricci scalar and trace of the energy-momentum tensor \cite{3g}. The astrophysical and cosmic indications of the $f(R,T)$ theory of gravity have been studied by many researchers \cite{4g}. However, the $f(R,T)$ or $f(R,L_m)$ theories can not be considered as the most generic Lagrangians for understanding the NMC between matter and geometry. One can generalize the above modified theories by invoking a term $R_{\alpha\beta}T^{\alpha\beta}$ in the Lagrangian. Some related work of such couplings can be seen in the Einstein Born-Infeld theories \cite{5g} where the role of square root in the Lagrangian is examined. It is worthy to stress that the addition of term $R_{\alpha\beta}T^{\alpha\beta}$ could provide NMC to the field at those environments where $f(R,T)$ fails to describe. For instance, in the scenario of trace-free stress energy tensor ($T=0$), the equations of motion of massive test particle can not entail NMC effects in $f(R,T)$ theory. However, $f(R,T,Q)$ theory ($Q\equiv R_{\alpha\beta}T^{\alpha\beta}$) can describe NMC in this context. Besides not spoiling the successful predictions of Solar systems, this theory has been found to stable against Dolgov-Kawasaki instability \cite{7}.

In this work, we have considered an extension of $f(R,T)$ gravity by taking into account a possible NMC between the energy-momentum tensor and the Ricci scalar. We characterized the gravitational field to produce a Lagrangian of the form $f(R,T,Q)$ (as used by Odintsov and S\'{a}ez-G\'{o}mez \cite{8}, for different parameters), where $f$ is an arbitrary function in the arguments $R,~T$, and $Q$, respectively. We derived the gravitational field equations for $f(R,T,Q)$ theory for the spherically symmetric spacetime which could provide an effective Einstein field equation. Generally, in $f(R,T,Q)$ theory, the concerned matter energy momentum tensor is non conserved and this phenomena leads to determine the additional force term which affects the motion of massive particles in gravitational field.

For the formulation of $f(R,T,Q)$ theory, we take an action \cite{21}
\begin{equation}\label{10}
I_{f(R,T,Q)}=\frac{1} {2}\int d^4x\sqrt{-g}
[f(R,T,R_{\alpha\beta}T^{\alpha\beta})+ \textit{L}_m],
\end{equation}
where, $R_{\alpha\beta}T^{\alpha\beta}$ and $T,~R$ indicate the
contraction of $T^{\alpha\beta}$ with Ricci tensor,
trace of stress energy tensor and Ricci
scalar, respectively. The link between
$T^{\alpha\beta}$ and matter lagrangian $\textit{L}_m$ is in indicated as follows
\begin{align}\label{11}
&T_{\alpha\beta}^{(m)}=-\frac{2}{\sqrt{-g}}\frac{\delta(\sqrt{-g}\textit{L}_m)}
{\delta{g^{\alpha\beta}}}.
\end{align}
The variation of (\ref{1}) with $g_{\alpha\beta}$, provides
\begin{align}\nonumber
&-G_{\alpha\beta}(f_{Q}\textit{L}_m - f_R) - g_{\alpha\beta}
\left\{\frac{f} {2}-\Box f_R
-\frac{R}{2}f_R-\frac{1}{2}\nabla_\pi\nabla_\rho(f_{Q}T^{\pi\rho})
-\textit{L}_mf_T\right\}\\\nonumber
&+2f_QR_{\pi(\alpha}T_{\beta)}^{~\pi}+\frac{1}{2}\Box(f_QT_{\alpha\beta})-\nabla_\pi\nabla_{(\alpha}
[T^\pi_{~\beta)}f_Q]-2\left(f_Tg^{\pi\rho}+f_QR^{\pi\rho}\right)\frac{\partial^2\textit{L}_m}{\partial
g^{\alpha\beta}\partial g^{\pi\rho}}\\\label{12} &-
T_{\alpha\beta}^{(m)}(f_T+\frac{R}{2}f_Q+1)-\nabla_\alpha\nabla_\beta
f_R=0,
\end{align}
here $\nabla_\pi,~\Box=g^{\alpha\beta}\nabla_\alpha\nabla_\beta,~G_{\alpha\beta}$ represent the
covariant derivative, d'Alembert's operator and Einstein tensor, respectively.
By fixing
$f(R,T,Q)= f(R,T)$ in Eq.(\ref{11}), the dynamics of $f(R,T)$ gravity can be resumed. One can get following expression from Eq.\eqref{12} as
\begin{align}\nonumber
&3\Box
f_R+\frac{1}{2}\Box(f_QT)-T(f_T+1)+\nabla_\pi\nabla_\rho(f_QT^{\pi\rho})+
R(f_R-\frac{T}{2}f_Q)\\\nonumber &+(Rf_Q+4f_T)\textit{L}_m
-2f+2R_{\pi\rho}T^{\pi\rho}f_Q -2
\frac{\partial^2\textit{L}_m}{\partial g^{\alpha\beta}\partial
g^{\pi\rho}}\left(f_Tg^{\pi\rho}+f_QR^{\pi\rho}\right).
\end{align}
During the motion of the particles in the gravitational field of $f(R,T,Q)$ gravity theory, an extra-force acts on particles due to which the energy momentum tensor is
non conservative,\cite{22} however, under some particular conditions, the
conservation can be obtained. It is well-known that generally $\textsl{L}_m$ has order of
two (or greater) in the metric tensor $g_{\alpha\beta}$. This means that its
second variation will be non zero. In the background of $f(R,T,R_{\mu\nu}T^{\mu\nu})$, Eq.(\ref{12}) can be rewritten as
\begin{equation}\label{13}
G_{\alpha\beta}=\kappa T^{\textrm{eff}}_{\alpha\beta},
\end{equation}
where
\begin{align}\nonumber
{{T}^{\textrm{eff}}_{\alpha\beta}}&=\frac{1}{(f_R-f_Q\textit{L}_m)}\left
[(f_T+\frac{1}{2}Rf_Q+1)T^{(m)}_{\alpha\beta}+
\left\{\frac{R}{2}\left(\frac{f}{R}-f_R\right)-\textit{L}_mf_T-\frac{1}{2}\right.\right.\\\nonumber
&\left.\left.\times\nabla_{\mu}\nabla_{\nu}(f_QT^{\mu\nu})\right\}g_{\alpha\beta}
-\frac{1}{2}\Box(f_QT_{\alpha\beta})
-(g_{\alpha\beta}\Box-\nabla_\alpha\nabla_\beta)f_R-2f_QR_{\mu(\alpha}T^\mu_{~\beta)}\right.\\\label{14}
&\left.+\nabla_\mu\nabla_{(\alpha}
[T^\mu_{~\beta)}f_Q]+2\left(f_QR^{\mu\nu}+f_Tg^{\mu\nu}\right)\frac{\partial^2\textit{L}_m}{\partial
g^{\alpha\beta}\partial g^{\mu\nu}}\right].
\end{align}
The non zero modified field equations for our spherical metric are
\begin{align}\nonumber
\mu^{\textrm{eff}}&=\frac{1}{(f_R+f_Q\mu)}\left[-\mu
f_T+\frac{f''_R}{B^2}+\left(\frac{2C'}{C}-\frac{B'}{B}\right)
\frac{f'_R}{B^2}-\left(\frac{\dot{B}}{B}+\frac{2\dot{C}}{C}\right)
\frac{\dot{f_R}}{A^2}+\mu\chi_1\right.\\\nonumber
&\left.-\frac{R}{2}\left(\frac{f}{R}-f_R\right)+\chi_2\dot{\mu}
+\chi_3\mu'+\frac{f_Q}{2A^2}\ddot{\mu}+\frac{f_Q}{2B^2}\mu''+\chi_4P_r
+\left(\frac{f'_Q}{B^2}-\frac{5}{2}f_QB'\right)\right.\\\label{15}
&\left.\times
P'_r+\frac{f_Q}{2B^2}P''_r-\frac{f_Q}{2A^2B}\dot{B}\dot{P_r}+\chi_5P_\bot-\frac{3f_Q}{2A^2C}\dot{P_\bot}\dot{C}
-\frac{3f_Q}{2B^2C}{P'_\bot}{C'}\right],\\\nonumber
P_r^{\textrm{eff}}&=\frac{1}{(f_R+f_Q\mu)}\left[\mu
f_T+\frac{\ddot{f_R}}{A^2}+\left(\frac{2\dot{C}}{C}-\frac{\dot{A}}{A}\right)\frac{\dot{f_R}}{A^2}-\left(\frac{A'}{A}
+\frac{2C'}{C}\right)\frac{f'_R}{B^2}+\chi_6P_r\right.\\\nonumber
&\left.+\frac{R}{2}\left(\frac{f}{R}-f_R\right)+\chi_7P'_r
+\chi_8\dot{P_r}+\chi_9P_\bot+\frac{f_Q}{A^2}\frac{\dot{C}}{C}\dot{P_\bot}
-\frac{f_Q}{2A^2}\ddot{\mu}+\frac{f_Q}{2B^2}\frac{A'}{A}\mu'\right.\\\label{16}
&\left.+\chi_{10}\mu+\chi_{11}\dot{\mu}+\frac{f_Q}{B^2}\frac{C'}{C}P'_\bot\right],\\\nonumber
P_\bot^{\textrm{eff}}&=\frac{1}{(f_R+f_Q\mu)}\left[\mu
f_T+\frac{\ddot{f_R}}{A^2}-\frac{f''_R}{B^2}+\left(\frac{\dot{B}}{B}-\frac{\dot{A}}{A}+\frac{\dot{C}}{C}\right)\frac{\dot{f_R}}{A^2}
+\chi_{12}P_\bot+\chi_{13}\dot{P_\bot}\right.\\\nonumber
&\left.+\chi_{14}P'_\bot+\frac{R}{2}\left(\frac{f}{R}-f_R\right)+\left(\frac{B'}{B}-\frac{A'}{A}
-\frac{C'}{C}\right)\frac{f'_R}{B^2}-\frac{f_Q}{2A^2}\ddot{P_\bot}+\frac{f_Q}{2B^2}{P''_\bot}\right.\\\nonumber
&\left.+\chi_{16}\mu+\frac{f_Q}{2A^2}\frac{\dot{B}}{B}\dot{P_r}+\frac{5f_Q}{2B^2}\frac{B'}{B}P'_r-\frac{P'_rf'_Q}{B^2}-\frac{f_Q}{2B^2}P''_r
+\left(\frac{5f_Q\dot{A}}{2A^3}-\frac{\dot{f_Q}}{A^2}\right)\dot{\mu}\right.\\\label{17}
&\left.-\frac{f_Q}{2A^2}\ddot{\mu}+\frac{f_Q}{2B^2}\frac{A'}{A}\mu'+\chi_{15}P_r\right],\\\nonumber
T_{01}^{\textrm{eff}}&=\frac{1}{(f_R+f_Q\mu)}\left[\dot{f'_R}
-\frac{A'}{A}\dot{f_R}-\frac{\dot{B}}{B}f'_R
\dot{\mu}'-\frac{\dot{\mu}A'}{A}
-\frac{\mu'\dot{B}}{B}+\dot{P_r}'-\frac{\dot{P_r}A'}{A}-\frac{\dot{B}P_r'}{B}\right.\\\nonumber
&\left.+q\left[AB(1+f_T)-\frac{\ddot{B}f_Q}{A}+\frac{3\dot{A}\dot{B}f_Q}{2A^2}
+\frac{3A''f_Q}{2B}+\frac{3A'B'f_Q}{2B^2}\right]
-\left[\frac{1}{2}\left(\frac{qB}{A}\right)^{\dot{}}\left(\frac{A'}{A}\right)f_Q\right.\right.\\\nonumber
&\left.\left.+\frac{1}{2}\left(\frac{qA}{B}\right)^{\dot{}}\left(\frac{B\dot{B}}{A^2}\right)f_Q\right]
-\left[\frac{1}{2}\left(\frac{qB}{A}\right)'\left(\frac{AA'}{B^2}\right)f_Q-\left(\frac{qA}{B}\right)'
\left(\frac{B'}{B}\right)
f_Q\right]\right.\\\label{18}&\left.+\frac{1}{2}\left(\frac{qB}{A}\right)^{\ddot{}}
+\frac{1}{2}\left(\frac{qA}{B}\right)''
\right]\equiv q^{eff}.
\end{align}
The values of $\chi_{i}$'s are linked to the space-time of spherically
symmetry and given in Appendix A.
One can write the following expression through Misner and Sharp formalism for the spherical geometry as
\begin{equation}\label{19}
m(r,t)=\frac{C}{2}\left[\frac{\dot{C^{2}}}{A^{2}}-\frac{C'^{2}}{B^{2}}+1\right].
\end{equation}
It is worthy to mention that this construct is defined in GR relativity and not in other theories of gravity. For
example, the generic form of above expression contains Newton's constant $G$, which becomes an effective
one (a field) $G_{eff}$ in $f(R,T,Q)$ gravity, so one can not say that the above expression could
describe the mass-energy as in GR.

\section{Dynamical Equation}

The contracted form of Bianchi identities in the background of particular modified gravity theory leads to
the following non-vanishing equations
\begin{align}\label{20}
&\left[\dot{\mu}^{\textrm{eff}}+\left(P_r^{\textrm{eff}}
+\mu^{\textrm{eff}}\right)\frac{\dot{B}}{B}+2(\mu^{\textrm{eff}}
+P_\bot^{\textrm{eff}})\frac{\dot{C}}{C}\right]\frac{1}{A}
+A{q'^{\textrm{eff}}}+Aq^{\textrm{eff}}\left(\frac{3A'}{A}
+\frac{B'}{B}+\frac{2C'}{C}\right)+Z_{3}=0,\\\label{21}
&\left[P'^{\textrm{eff}}_r+\left(P_r^{\textrm{eff}}
+\mu^{\textrm{eff}}\right)\frac{A'}{A}+2\left(P^{\textrm{eff}}_r
-P_\bot^{\textrm{eff}}\right)\frac{{C'}}{C}\right]\frac{1}{B}+B{\dot{q}^{\textrm{eff}}}+Bq^{\textrm{eff}}
\left(\frac{\dot{A}}{A}+\frac{3\dot{B}}{B}+\frac{2\dot{C}}{C}
\right)+Z_{4}=0.
\end{align}
The values of $Z_3$ and $Z_4$ are given in the Appendix A.
Here, the notation `eff' indicating the influence of $f(R,T,Q)$ gravity within the concerned matter quantities.
To examine the dynamical conditions on the system, we define a differential operator for
proper time as $D_{T}$ given by
\begin{equation}\label{22}
D_{T}=\frac{1}{A}{\frac{\partial}{\partial{t}}},
\end{equation}
and correspondingly the proper radial differential operator $D_{R}$
as
\begin{equation}\label{23}
D_{R}=\frac{1}{\acute{R}}\frac{\partial}{\partial{r}},
\end{equation}
while $R=C$ is the proper radius inside the spherical surface $\Sigma$.
For the collapsing matter configuration, we can define the velocity $U$ which is the rate of
proper radius and proper time as $U=rD_{T}C<0$ so that Eq.(\ref{18}) takes the form
\begin{equation}\label{24}
E=\frac{\acute{R}}{B}.
\end{equation}
By making use of Eqs.(\ref{13})-(\ref{16}) and (\ref{21})-(\ref{23}), we determine
\begin{equation}\label{25}
 D_{T}m=-\frac{\dot{R}R^{2}P^{\textrm{eff}}_r}{2A}+\frac{\acute{R}R^{2}q^{\textrm{eff}}}{2AB^{2}},
\end{equation}
and
\begin{equation}\label{26}
D_{R}m=\frac{R^{2}\mu^{\textrm{eff}}}{2}-\frac{\dot{R}R^{2}q^{\textrm{eff}}}{2\acute{R}A^{2}}.
\end{equation}
Equation (26) demonstrate the inner variational
rate of complete energy of the surface having radius $R=C$. The term $P_{r}^{\textrm{eff}}$ describes that when $U<0$ the collapse occurs while, the amount of energy increases inside the radius $r$ is the work done by $q^{\textrm{eff}}$. Since, pressure is zero in the diffusion process so the flow of heat flux is negligible.
Since, we are assuming the center of distribution as regular so, $m(0)=0$ and the above equation can be integrated to obtain
\begin{equation}\label{27}
m=\int\left(\frac{R^{2}\mu^{\textmd{eff}}}{2}-\frac{\dot{R}R^{2}q^{\textmd{eff}}}{2{R'}A^{2}}\right)dR.
\end{equation}
To get acceleration of the astronomical elementary matter under the boundary $\sum$, we used the Eq (\ref{17}),(\ref{19}),(\ref{22}), and (\ref{24}), and obtain
\begin{equation}\label{28}
D_{T}U=-\frac{R}{2}{P^{\textmd{eff}}_{r}}-\frac{m}{R^{2}}+\frac{E{A'}}{AB}
\end{equation}
and then substituting the value of $\frac{\acute{A}}{A}$ from Eqs.(\ref{19})and (\ref{20}), we get
\begin{eqnarray}\nonumber
\left(P^{\textmd{eff}}_{r}+\mu^{\textmd{eff}}\right)D_{T}U&=&
-\left(P^{\textmd{eff}}_{r}+\mu^{\textmd{\textmd{eff}}}\right)\left[\frac{m}{R^{2}}
-\frac{RP^{\textmd{eff}}_{r}}{2}\right]\\\nonumber
&-&E^{2}\left[\frac{2}{R}\left(P^{\textmd{eff}}_{r}-P_\bot^{\textmd{eff}}\right)
+D_{R}{R'^{\textmd{eff}}_{r}}\right]\\\label{29}
&-&E\left[ABD_{T}q^{\textmd{eff}}+Bq^{\textmd{eff}}\left(\frac{\dot{A}}{A}
+\frac{2\dot{C}}{C}+\frac{3\dot{B}}{B}\right)\right]Z_{4}.
\end{eqnarray}
The right side of the above equation have contribution of different forces. The factor in the bracket (as similar to the term in left side) tell us the material  mass density (which is also known as ``passive" density of gravitational mass).We can see the contribution of two different terms with the contribution of $(R,T,Q)$ gravity in the square bracket. The first factor is inclination of the complete  effective pressure which is radial in nature. The second term shows the locally anisotropic pressure and influence of $P_{r}$ in the background of $f(R,T,Q)$ gravity\cite{23}. The factor which is in the second square brackets represents the dissipation effect under the influence of $f(R,T,Q)$ gravity, known as ``active" factor of gravitational mass.

\section{The Transport Equation}

From the phenomenological theory formulated by Muller-Israel-Stewart for the dissipation of fluids, we can utilize the heat transport equation \cite{24}. It is well-known that parabolic equation which is derived by the Maxwell-Fourier law, indicates the flow of heat with infinite speed \cite{25,26,27}. For the dissipative process in general relativity, this is the central point of pathologies \cite{28} as established by Eckart and Landau \cite{29}. Many theories have been intended in the past to overthrow these difficulties with persisting relaxation times. The main point is that all suggested theories \cite{30} are about heat transport equation, not following the Maxwell-Fourier law, which gives us a hyperbolic equation in the process of perturbation. The transport equation under diffusion process can be written as
\begin{equation}\label{30}
\tau h^{\alpha\beta}V^{\gamma}q_{\beta;\gamma}+q^{\alpha}=-\kappa
h^{\alpha\beta}(T_{,\beta}+Ta_{\beta})-\frac{\kappa
T^{2}}{2}\left(\frac{\tau V^{\beta}}{\kappa T^{2}}\right)_{;\beta} q^{\alpha},
\end{equation}
where $h^{\alpha\beta}$ is an orthogonal projector to $V^\mu$ in three space, $\kappa$ indicates the thermal conductivity, $T$ is the temperature while $\tau$
denotes the relaxation time. There exists only one independent component of Eq.(\ref{31}) due to the symmetry of problem, which by using Eqs.(\ref{1}), (\ref{6}) and (\ref{10}) turns out to be
\begin{equation}\label{31}
\tau\dot{q^{\textmd{eff}}}=-\frac{\kappa}{B}\acute{TA}\tau
q^{\textmd{eff}}\left(\frac{\dot{B}}{2B}+\frac{\dot{C}}{C}\right)-\frac{1}{2}\kappa
q^{\textmd{eff}}T^{2}\dot{\frac{\tau}{\kappa T^{2}}}-q^{\textmd{eff}}A.
\end{equation}
By making use of Eqs.(\ref{21})-(\ref{25}), we obtain
\begin{eqnarray}\nonumber
D_{T}q^{\textmd{eff}}&=& -\frac{1}{2}\kappa
T^{2}q^{\textmd{eff}}D_{T}\left(\frac{\tau}{\kappa
T^{2}}\right)-q^{\textmd{eff}}\left(\frac{3U}{2R}+\frac{\sigma}{2}+\frac{1}{\tau}\right)-\frac{\kappa
E}{\tau}D_{R}T-\frac{\kappa T}{\tau E}D_{T}U\\\label{32}
&-&\frac{\kappa T}{\tau
E}\left(\frac{1}{R^{2}}(m-\frac{R^{3}}{2}P^{\textmd{eff}}_{r})\right).
\end{eqnarray}
Now, we can couple the above heat transport equation with our dynamical equation as obtained in Eq.(\ref{30}) with some manipulations to get
\begin{eqnarray}\nonumber
\left(P^{\textmd{eff}}_{r}+\mu^{\textmd{eff}}\right)(1-\alpha)D_{T}U &=&
(1-\alpha)F_{grav}+F_{hyd}+\frac{\kappa
 E^{2}}{\tau}D_{R}T+E\left[\frac{\kappa AT^{2}}{2\tau}D_{T}(\frac{\tau}{\kappa
 T^{2}})\right]\\\label{33}
&-&Eq^{\textmd{eff}}\left[-\frac{3U}{2R}-\frac{1}{2}\sigma-\frac{1}{\tau}
 +\left(\frac{\dot{A}}{A}+\frac{2\dot{C}}{C}+\frac{2\dot{B}}{B}\right)\right]Z_{4},
\end{eqnarray}
where ${\alpha}$ is given by
\begin{equation}\label{34}
\alpha=\frac{\kappa T}{\tau(P^{\textmd{eff}}_{r}+\mu^{\textmd{eff}})},
\end{equation}
while $F_{grav},~F_{hyd}$ are
\begin{eqnarray}\nonumber
F_{grav}&=&\left(P^{\textmd{eff}}_{r}+\mu^{\textmd{eff}}\right)
\left(\frac{1}{R^{2}}(m-\frac{R^{3}}{2}P^{\textmd{eff}}_{r})\right),\\\nonumber
F_{hyd}&=&-E^{2}\left(D_{R}P^{\textmd{eff}}_{r}+\frac{2}{R}(R^{\textmd{eff}}_{r}-P_\bot^{\textmd{eff}})\right).
\end{eqnarray}
We have observed that under the influence of $f(R,T,Q)$ gravity the density of internal energy and density of passive gravitational mass, i.e., the term which is multiplying with
$D_{T}$ and the term in the right side of Eq.(\ref{33}) is decreased due to the factor $(1-\alpha)$ in the radiative phenomena. It is worthy to mention here that a similar factor $(1-\alpha)$ has been
found by Herrera et al. \cite{jz1} with spherical geometry in the presence of heat flux and anisotropic pressure
within the context of GR, due to which the density of internal energy
and density of passive gravitational mass is decreased, but here
generalized with the inclusion of $f(R,T,Q)$
gravity. Herrera et al. \cite{jz2} also presented a more general treatment to discuss the
dynamics of viscous dissipative gravitational collapse
which affect the density of internal energy
and density of passive gravitational mass by including bulk viscosity, shear viscosity and
radiation density with spherical symmetry in the context of GR. They also compared the
obtained results with the previous works where the viscosity and radiation terms were neglected
in the framework of GR.

\section{The Weyl tensor}

Here, we explored an important explicit relationship between the Weyl tensor and matter variables, which give us some concluding results about the evolution of time\cite{31}. From Weyl tensor,
we formulate Weyl scalar $\mathcal{C}^{2}=C^{\alpha\beta\gamma\delta}C_{\alpha\beta\gamma\delta}$ which gives the value in the form
of Kretschmann scalar $\textit{R}=R^{\alpha\beta\gamma\delta}R_{\alpha\beta\gamma\delta}$, Ricci tensor $R_{\alpha\beta}$ and the curvature scaler $R$ as
\begin{equation}\label{35}
\mathcal{C}^{2}=\textit{R}-2R^{\alpha\beta\gamma\delta}R_{\alpha\beta\gamma\delta}+\frac{1}{3}R^{2}.
\end{equation}
Now, by substituting the value of Kretschmann scalar and using the field equations, we get
\begin{equation}\label{36}
\varepsilon=m-\frac{1}{6}\left(\mu^{\textmd{eff}}-P^{\textmd{eff}}_r+P_\bot^{\textmd{eff}}\right),
\end{equation}
here
\begin{equation}\label{37}
\varepsilon=\frac{\mathcal{C}}{48^{1/2}}R^{3},
\end{equation}
Now, making use of Eqs.(\ref{26}), (\ref{27}) and (\ref{37}), we get
\begin{equation}\label{38}
D_{T}\varepsilon=-\frac{1}{6}D_{T}\left(\mu^{\textmd{eff}}-P_r^{\textmd{eff}}
+P_{\bot}^{\textmd{eff}}\right)-\frac{\dot{R}R^{2}P^{\textmd{eff}}_r}{2A}
+\frac{\acute{R}R^{2}q^{\textmd{eff}}}{2AB^{2}},
\end{equation}
and
\begin{equation}\label{39}
D_{R}\varepsilon=-\frac{1}{6}D_{R}\left(\mu^{\textmd{eff}}
-P_r^{\textmd{eff}}+P_{\bot}^{\textmd{eff}}\right)+\frac{R^{2}\mu^{\textmd{eff}}}{2}
-\frac{\dot{R}R^{2}q^{\textmd{eff}}}{2\acute{R}A^{2}},
\end{equation}
From Eq.(\ref{39}), we find the expression for perfect fluid which is non-dissipative and non-charged in nature as,
\begin{equation}\label{40}
D_{R}\varepsilon+\frac{1}{6}D_{R}\mu^{eff}=0.
\end{equation}

If we set $D_{R}\mu^{eff}=0$, then it further brings out $\mathcal{C}=0$ which is formulated by using the condition of regular axis, and if we use the conformally flat condition, we get the homogeneous energy density. For the perfect fluid, this specific relation in Weyl tensor and the inhomogeneous nature of energy density
is proposed by Penrose, providing the significance of gravitational arrow of time involving Weyl tensor \cite{32}. The phenomena behind this proposal includes the basic properties of tidal forces, i.e.,
as the evolution occurs the inhomogeneity in the gravitating fluid increases. However, in the presence of dissipation or local anisotropy, such relationship is not valid.

\section{Conclusion}

Self-gravitating fluids have attracted many researchers due to their
applications in relativistic astrophysics. We discussed the gravitational collapse of self-gravitating spherically symmetric star by including the dissipation effects in the matter distribution under the background of particular modified gravity theory. We calculated all the parametric equations in order to observe the dissipative effects on varying energy density and explored modified field equations with anisotropic matter distribution. The conservation equations are also explored
to examine the dynamical equation that enable us to study the inner variational rate of complete energy and the variation of confined energy between spherical surfaces. The mass function is defined in order to achieve the acceleration of astronomical particles inside boundary $\sum$ and noticed the effect of gravitational force in $f(R,T,Q)$ gravity. We explored the heat transport equation with the effective terms to get the dissipation effects, and then coupled it with the dynamical equation and discussed the phenomena of gravitational collapse. The main findings can be summarized as:
\begin{itemize}
\item It turns out that the coupled dynamical heat transport equation
has an extra factor, $\alpha$, due to heat flux as obtained in Eq.(\ref{34}). We observe that the extra curvature ingredients due to modified gravity does not enter into the term $(1-\alpha)$ but has influence on the gravitational mass indicating how thermal effects reduce the effective inertial mass.
Consequently, both the internal energy density and density of gravitational mass vanishes.
It indicates that inertial force is absent there and matter would
observe a gravitational attraction leading to collapse.
\item When $0<\alpha<1$, the energy mass density gradually decreasing.
\item When $\alpha>1$, it shows an increase in energy mass density.
According to equivalence principle, a decrease or increase of mass should occur in a dynamical system so that one can discriminate between collapsing and expanding process during the evolution.
\end{itemize}
A local anisotropic pressure and the dissipative process is involved with the contribution of $f(R,T,Q)$ gravity.
Considering the imploding object evolving in such a manner that $\alpha$ may increases its impact and ends upto $1$
for some region. The gravitational force term
decreases in this process leading to a change of the sign of the right hand side
of Eq.(\ref{34}). This would happen for small values of the
effective inertial mass density and implies a strong bouncing of
that part of the sphere. We noticed that
the term $(1-\alpha)$ (in the inertial mass,
gravitational force) is related to the left of
Eq.(\ref{34}). We have also explored a relation linking the
Weyl tensor and the inhomogeneous nature of density.

\section*{Acknowledgments}

The work of KB was supported in part by the JSPS KAKENHI Grant Number JP
25800136 and Competitive Research Funds for Fukushima University Faculty
(18RI009). The works of ZY and MZB have been supported financially by National
Research Project for Universities (NRPU), Higher Education Commission Pakistan under research
Project No. 8769.

\vspace{0.3cm}

\section{Appendix}

The values of $\chi_i$'s in the field equations are
\begin{align}\nonumber
\chi_1&=1+f_T-\frac{3R}{2}f_Q-\frac{\dot{A}\dot{f_Q}}{2A^3}
+\frac{4\dot{A}^2}{A^4}f_Q-\frac{7\dot{A}}{2A^3}\dot{f_Q}
+\frac{2A'f'_Q}{AB^2}+\frac{A'^2f_Q}{A^2B^2}+\frac{A''f_Q}{AB^2}\\\nonumber
&-\frac{\dot{B}\dot{f_Q}}{2A^2B}-\frac{\dot{A}\dot{B}}{A^3B}f_Q
-\frac{B'f'_Q}{2B^3}-\frac{A'B'}{AB^3}f_Q
-\frac{\dot{C}\dot{f_Q}}{A^2C}-\frac{2\dot{A}\dot{C}}{A^3C}f_Q
+\frac{C'f'_Q}{CB^2}+\frac{2A'C'}{AB^2C}f_Q,\\\nonumber
\chi_2&=-\frac{11}{2A^3}f_Q\dot{A}-\frac{f_Q\dot{B}}{2A^2B}
-\frac{f_Q\dot{C}}{A^2C},~
\chi_3=\frac{f_Q}{CB^2}C'+\frac{f'_Q}{B^2}
-\frac{2A'f_Q}{AB^2}-\frac{B'f_Q}{2B^3},\\\nonumber
\chi_4&=\frac{f''_Q}{2B^2}+\frac{4B'^2f_Q}{B^4}
-\frac{B''}{B^3}f_Q-\frac{5B'f'_Q}{2B^3}-\frac{\dot{B}\dot{f_Q}}{2A^2B}
+\frac{\dot{B}^2f_Q}{A^2B^2},\\\nonumber
\chi_5&=\frac{3\dot{C}^2f_Q}{A^2C^2}-\frac{3\dot{C}\dot{f_Q}}{2A^2C}
-\frac{3C'f'_Q}{2B^2C}+\frac{3C'^2f_Q}{B^2C^2},\\\nonumber
\chi_6&=1+f_T-\frac{3}{2}Rf_Q-\frac{3B'^2f_Q}{B^4}
+\frac{2B''}{B^3}f_Q+\frac{B'}{B^3}f'_Q-\frac{5\dot{B}\dot{f_Q}}{2A^2B}
-\frac{3\dot{B}^2f_Q}{A^2B^2}\\\nonumber
&-\frac{B\ddot{B}}{A^2B^2}f_Q-\frac{\ddot{f_Q}}{2A^2}
+\frac{\dot{A}\dot{f_Q}}{2A^3}+\frac{\dot{A}\dot{B}}{A^3B}f_Q+\frac{A'f'_Q}{2AB^2},\\\nonumber
\chi_7&=\frac{4B'}{B^3}f_Q+\frac{A'}{2AB^2}f_Q,~\chi_8
=\frac{\dot{B}}{2A^2B}f_Q+\frac{\dot{A}f_Q}{2A^3}-\frac{5\dot{B}f_Q}{2A^2B},\\\nonumber
\chi_9&=\frac{\dot{C}\dot{f_R}}{A^2C}
-\frac{2\dot{C}}{A^2C^2}f_Q,~\chi_{11}=\frac{5\dot{A}f_Q}{2A^3}-\frac{\dot{f_Q}}{A^2},\\\nonumber
\chi_{10}&=\frac{5\dot{A}\dot{f_Q}}{2A^3}-4\frac{\dot{A}^2}{A^4}f_Q+\frac{\ddot{A}}{A^3}f_Q-\frac{\ddot{f_Q}}{2A^2}+\frac{A'f'_Q}{2AB^2}-\frac{A'^2f_Q}{A^2B^2},\\\nonumber
\chi_{12}&=1+f_T-\frac{3}{2}Rf_Q-\frac{5\dot{C}^2}
{A^2C^2}f_Q+\frac{2C'}{CB^2}f_Q-\frac{2\dot{C}\dot{f_Q}}{A^2C}
-\frac{\ddot{C}f_Q}{A^2C}
-\frac{\ddot{f_Q}}{2A^2}+\frac{\dot{A}\dot{f_Q}}{2A^3}\\\nonumber
&+\frac{\dot{A}\dot{C}}{CA^3}f_Q+\frac{A'f'_Q}{2AB^2}
+\frac{A'C'}{ACB^2}f_Q+\frac{2C'f'_Q}{B^2C}+\frac{C'^2f_Q}{B^2C^2}
+\frac{C''}{B^2C}f_Q
-\frac{\dot{B}\dot{f_Q}}{2A^2B}\\\nonumber
&+\frac{f''_Q}{2B^2}-\frac{\dot{C}\dot{B}}{CA^2B}f_Q
-\frac{B'f'_Q}{2B^3}-\frac{C'B'}{CB^3}f_Q,\\\nonumber
\chi_{13}&=\frac{\dot{C}\dot{A}}{2A^3C^2}f_Q
-\frac{2\dot{C}}{A^2C}f_Q-\frac{\dot{f_Q}}{A^2}-\frac{\dot{B}f_Q}{2A^2B},\\\nonumber
\chi_{14}&=\frac{2C'}{B^2C}f_Q+\frac{A'f_Q}{2AB^2}
+\frac{2C'f_Q}{B^2C}+\frac{f'_Q}{B^2}-\frac{B'f_Q}{2B^3},\\\nonumber
\chi_{15}&=\frac{5B'f'_Q}{2B^3}-\frac{4B'^2}{B^4}f_Q
+\frac{B''}{B^3}f_Q-\frac{f''_Q}{2B^2}
+\frac{\dot{B}\dot{f_Q}}{2A^2B}-\frac{\dot{B}^2f_Q}{A^2B^2},\\\nonumber
\chi_{16}&=\frac{5\dot{A}}{2A^3}\dot{f_Q}
-\frac{4\dot{A}^2}{A^4}f_Q
+\frac{\ddot{A}}{A^3}f_Q-\frac{\ddot{f_Q}}{2A^2}
-\frac{A'^2f_Q}{A^2B^2}+\frac{A'f'_Q}{2AB^2}.
\end{align}

The values of $Z_3$ and $Z_4$ in the conservation equations are
\begin{align}\nonumber
Z_{3}&=\frac{2}{1+R f_{RT}+2f_T}\left[2\dot({\mu}{f_T})
-\dot({f_{RT}(R^{00})\mu})+(P_r f_{RT} R^{10})'
-\frac{P_\bot}{C^2}\left(1+\frac{1}{\sin^2\phi}
+\frac{\mu}{A^2}+\frac{P_r}{B^2}\right)_{,0}\right.\\\nonumber&\times\left.{f_{RT}\sum R_{ii}+f_T(-A^2+B^2+C^2+C^2\sin^2\phi)}
-G^{00}\dot{\mu f_T}-G^{10}(\mu f_T)'-\frac{\mu}{2 A^2}{(\dot{R f_{RT}})+2\dot{f_T}}\right],\\\nonumber
Z_4&=\frac{2}{1+R(f_{RT})+2f_T}\left[2(\mu(f_T))'-\dot({(f_{RT})R^{01}(\mu))}+(P_r(f_{RT})R^{11})'\right.
-\left(\frac{P_\bot}{C^2}\left(1+\frac{1}{\sin^2\phi}\right.\right.\\\nonumber
&\left.\left.+\frac{\mu}{A^2}+\frac{P_r}{B^2}\right)\right)'
{f_{RT}\sum R_{ii}+f_T(-A^2+B^2+C^2+C^2\sin^2\phi)}
-G^{01}\dot{(\mu f_T)}-G^{11}(\mu f_T)'\\\nonumber&-\left.\frac{P_r}{2A^2}{(R f_{RT})'+2f_T}\right].
\end{align}

\renewcommand{\theequation}{A\arabic{equation}}
\setcounter{equation}{0}
\end{document}